\documentstyle[preprint,aps]{revtex}
\begin{document}
\title{Monte Carlo calculation of the current-voltage
characteristics of a two dimensional lattice Coulomb gas}
\author{Hans Weber{$^1$}, Mats Wallin{$^2$}, 
and Henrik Jeldtoft Jensen{$^3$}}
\address{{$^1$}Department of Physics, Lule{\aa} University of Technology,
S-971 87 Lule{\aa}, Sweden}
\address{{$^2$}Department of Theoretical Physics,
Royal Institute of Technology, S-100 44 Stockholm, Sweden}
\address{{$^3$}Department of Mathematics,
Imperial Collage, London SW7 2BZ, United Kingdom}
\maketitle

\begin{abstract}
We have studied the nonlinear current-voltage characteristic of
a two dimensional lattice Coulomb
gas by Monte Carlo simulation.
We present three different determinations of the
power-law exponent $a(T)$ of the nonlinear current-voltage
characteristic, $V \sim I^{a(T)+1}$.
The determinations rely on both equilibrium and 
non-equilibrium simulations.
We find good agreement between the different determinations,
and our results also agree closely with
experimental results for Hg-Xe thin film superconductors
and for certain single crystal thin-film high temperature superconductors.
\end{abstract}

\section{Introduction}

In two dimensions the superconducting transition
in zero magnetic field is a Kosterlitz-Thouless 
transition. \cite{kt,Bere,minnhagen}
This has been verified over the years in both
experiments \cite{kadin} and in many models of
superconductors like the XY, Villain,
and Coulomb gas models \cite{kt,teitel4,villain}.
The important degrees of freedom in a system undergoing
a Kosterlitz-Thouless transition are thermally excited vortex pairs.
The Kosterlitz-Thouless transition is sometimes
also referred to as a vortex unbinding transition,
as for temperatures below the transition temperature
$T_c$ all vortices are bound in neutral pairs. 
These pairs start to unbind at and above $T_c$.

A typical way to look for a Kosterlitz-Thouless transition
in experiments on thin superconducting films
is to probe the current-voltage (IV)
characteristic \cite{kadin,gorlova,pradhan}.
Both the linear and the nonlinear IV characteristics
have specific fingerprints identifying a Kosterlitz-Thouless transition.
Vortices determine the IV characteristic for the following reasons:
If a vortex is dragged across the system a voltage is induced.
Hence resistance is zero only if there are no vortices available
to move across the system,
and only then the system is truly superconducting.
Vortices that are bound in neutral pairs are unable to move freely
and to cause dissipation.
However an external applied in-plane 
supercurrent yields a perpendicular Lorentz force acting in opposite
direction on vortices with different vorticity.
This gives a net flux of vortices across the system,
which shows up as nonlinear (i.e.\ current dependent) resistance.

Below the Kosterlitz-Thouless transition temperature 
all vortices are bound in neutral pairs 
by the logarithmic vortex interaction,
and the linear resistance is thus zero.
Therefore the system superconducts below the Kosterlitz-Thouless transition.
The linear resistance drops to zero at the Kosterlitz-Thouless transition
with an exponential functional form,
$R \sim \xi^{-2}$ with $\ln \xi \sim |T-T_c|^{-1/2}$ \cite{minnhagen}.
This is consistent with experiments, although the
logarithm is a complication for quantitative comparison
between theory and experiment.
A finite applied current gives a power-law nonlinear IV characteristic
of the form
$V \sim I^{a(T) + 1}$.
The critical current is thus zero.
At the Kosterlitz-Thouless transition the IV exponent $a(T)$
assumes the universal value 2, so $V \sim I^3$ at $T=T_c$.
For $T<T_c$ one has
$a(T)>2$, and for $T>T_c$ one has $a(T)=0$
(for small enough currents) \cite{NK}.
Experiments on, for example, thin Hg-Xe alloy films \cite{kadin}
and also for certain single crystal 
high temperature superconductors \cite{gorlova,pradhan},
among some, have confirmed this.

Since IV characteristics are hard to calculate analytically
computer simulation is a useful tool.
IV characteristics of vortex systems have recently been 
calculated successfully with Monte Carlo simulations \cite{MWSMG}.
Linear and nonlinear IV characteristics of vortex glass superconductors
have been reported in Refs.~\onlinecite{MWSMG,Hyman}.
In a recent Monte Carlo simulation of the Coulomb gas the linear resistance
was used to locate the Kosterlitz-Thouless transition \cite{wallweb}.
The nonlinear IV characteristics at the Kosterlitz-Thouless transition 
has been calculated in Ref.~\onlinecite{teitel3},
and a finite-size scaling analysis accurately
verified the relation $V \sim I^3$ at the Kosterlitz-Thouless transition.

In this paper we study the IV characteristics of a lattice Coulomb gas model
by Monte Carlo simulations of vortex dynamics.
We calculate the IV exponent $a(T)$ of the
Coulomb gas in three different ways:
(1) By direct Monte Carlo calculation of the nonlinear resistance,
(2) by a self consistent linear screening construction for the
energy barrier for current induced vortex-pair breaking 
giving thermally activated resistance,
and (3) by a finite scaling construction from data for the linear resistance.
All methods are based on Monte Carlo simulations, and we apply both
equilibrium and non-equilibrium simulations.
These three methods give the same results, giving us a consistent
and simple picture of nonequilibrium response in this system.
Furthermore, we compare our results for $a(T)$ with
experiments. Scaling arguments give that $a(T)$ is a
universal scaling function of a reduced Coulomb gas temperature
$X = T/T_c$, and this is verified in experiments \cite{minnhagen}.
We find close agreement between our Monte Carlo results
and the experimental universal scaling curve,
and this comparison appears to be presented here for the first time.
The agreement between different methods, and between our 
simulations and experiments, are the main results of our paper.  
Some of our Monte Carlo results for the nonlinear IV characteristics have 
been obtained previously \cite{teitel3},
as explained above.

The paper is organized as follows: 
In Section II we define the lattice Coulomb gas model.
In Section III we study various approaches to the IV characteristics.   
In Section IV we describe our Monte Carlo methods for calculating IV characteristics.
In Section V we present the Monte Carlo results. 
Section VI contains discussion and conclusions.

\section[11]{Lattice Coulomb gas}

A useful starting point for calculations with superconductors in
the presense of currents and fields
is the Ginsburg-Landau model, with
the order parameter
$\Psi({\bf r}) = |\Psi({\bf r})| e^{i\phi({\bf r})}$
describing the 
superconducting order of the system.
However, this model  does not focus
particularly on vortex degrees of freedom.
The vortices constitute the essential degrees of freedom
near the Kosterlitz-Thouless transition.
An approximation to the Ginsburg-Landau model which
focuses only on the vortices is given by the Coulomb gas model.
Here thermal fluctuations in the magnitude of $\Psi$ are neglected,
since they are relevant only close to the mean-field 
transition temperature,
which is assumed to be well above the 
vortex transition temperature $T_c$.
In our simulations the model is discretized and put on
a lattice. The approximation made in the 
lattice discretization will only affect the short
range behavior of the vortices, as the lattice
defines the smallest possible separation.
The critical properties will however not be effected.
In general, large length scale properties should be 
reasonable modeled by the lattice Coulomb gas close to $T_c$.

The lattice Coulomb gas \cite{JKKN,teitel2} 
is defined by the partition
function $Z$ on a square lattice of side length $L$ using
periodic boundary conditions:
\begin{eqnarray}
\label{sysdef}
Z & = & \mbox{Tr}_n \exp[-\beta(H - \mu N)]  \\
\label{sysham}
H & = & \frac{1}{2} \sum_{i,j} n_i G_{ij} n_j  \\
N & = & \sum_{i} \mid n_i \mid ,
\end{eqnarray}
where $H$ is the Hamiltonian, 
$n_i$ is the vorticity at site $i$ (Coulomb gas charge),
$\mu=-E_c$ is the vortex
``chemical potential'' and $E_c$ is the vortex core energy,
and $T = 1/\beta$ is the Coulomb gas temperature \cite{minnhagen}.
The trace is over $n_i = 0, \pm 1$ 
on all sites $i$, subject to overall neutrality, $\sum_i n_i = 0$.
$G_{ij}$ is the lattice Green's function for the
logarithmic 2D vortex interaction,
\begin{equation}
G_{ij} = \frac{1}{L^2} \sum_{{\bf k}}
\frac{\pi e^{i \ {\bf k \cdot} ( {\bf r}_i - {\bf r}_j )}}
{2-\cos(k_x) - \cos(k_y)},
\label{green}
\end{equation}
where ${\bf k}$ are the reciprocal
lattice vectors, $k_x,k_y = 2\pi n/L, n=0,\dots,L-1$.

We will calculate the response voltage to an applied current
imposed on the Coulomb gas.
The above definition does not include any net currents.
How to include them and to calculate IV characteristics
by Monte Carlo simulation is described in the next section.

\section{Current-Voltage Characteristics}

In this section we discuss various aspects and approaches to the 
current-voltage characteristics of 2D superconductors close to the 
Kosterlitz-Thouless transition.

\subsection{Linear resistance}

A basic experiment on a superconductor is to measure
the linear resistance.
Such measurements on thin films of both 
conventional low-$T_c$ superconductors \cite{kadin} and 
single crystal high-$T_c$ materials \cite{gorlova,pradhan},
have been successfully interpreted
in terms of thermally excited  vortex fluctuations analyzed
by use of the Coulomb gas \cite{minnhagen}.

The linear resistivity is defined by 
$\rho = E/j$ for $j \rightarrow 0$,
where $j$ is the applied supercurrent density and $E$ is the
resulting induced electric field.
Some words about notation:
Since resistance and resistivity have the same 
dimension in two dimensions and our system is homogeneous,
they are the same, and they we will both be denoted by $R$.
$R$ will be reserved for linear resistance,
and will not be used to denote nonlinear resistance.
An applied supercurrent is denoted by $I = jL$,
and voltage is $V=EL$.

To determine the linear resistance in simulations of 
the Coulomb gas from $E/j$ for small $j$
has its limitations, as we have to
repeat the calculation at a sequence of 
current densities $j$, to make sure that $j$ is small
enough to be in the linear regime.
If the purpose is to measure only the linear resistance,
and not $E$ as function of $j$,
a different approach is to use
the Nyquist formula~\cite{Reif},
which relates the linear resistance to the equilibrium voltage fluctuations:
\begin{equation}
R = \frac{1}{2 T} \int_{-\infty}^{+\infty} dt\
\langle V(t) V(0) \rangle,
\label{Nyquist}
\end{equation}
where $V(t)$ is the induced voltage from
vortex motion at time $t$.
As an alternative to eq. \ref{Nyquist} the Kubo formula 
for the  vortex currents $I_v$, $R = \frac{1}{2T} \int_{-\infty}^{\infty}
dt\ \langle I_v(t) I_v(0) \rangle$
can be used. Given the Josephson relation we see 
immediately that the Kubo formula equals the Nyquist relation.


The linear resistance has been successfully used 
in a simulation  \cite{wallweb} to locate the
Kosterlitz-Thouless transition temperature $T_c$ of the 2D lattice Coulomb gas.
They find the finite size scaling relation at $T_c$:
\begin{equation}
L^2 R \left( 1 + \frac{1}{4 \ln(L) + C} \right) = constant  
\;\; \mbox{at} \;\; T=T_c,
\label{LogCorrection}
\end{equation}
to be valid to a very high precision.

The scaling relation
Eq. (\ref{LogCorrection}) was derived from
the following argument.
We assume the dynamical exponent $z=2$
for free vortex diffusion in two dimensions \cite{teitel3,AHNS}.
The linear resistance is a dynamical quantity,
it relates to the correlation time
$\tau$, which at $T_c$ diverges like $\tau \sim \xi^z$, where
$\xi$ is the correlation length.
According to the Josephson relation the voltage $V \sim
 \frac{d \Delta \phi}{d t} \sim \tau^{-1}$, where
$\Delta\phi$ is the gradient of the phase of the Ginsburg-Landau order
parameter \cite{tinkham}.
Therefore, we expect the linear resistance, Eq. (\ref{Nyquist}),
to scale like $R \sim \xi^{-2}$ at $T_c$.
At $T_c$ the correlation length diverges
and is cut of by the finite
size $L$ of the lattice and hence
$RL^2 = const$ at $T_c$, to lowest order.
The scaling relation has a logarithmic correction which has been included in 
Eq. (\ref{LogCorrection}).  This correction is readily obtained from the
corresponding correction terms for $1/\epsilon$ and
$\lambda$ \cite{weber}.

\subsection{Thermally activated resistance}

The above scaling argument led to a finite size scaling formula
which is useful for locating the transition temperature
from Monte Carlo data for the resistance of finite samples.
Here we will do a more detailed analysis that will also lead to
the same formula.
The analysis here does not directly involve scaling arguments,
but considers the interactions between vortices in the Coulomb gas.  
The analysis will give expressions for 
the resistance from thermally activated free vortices in
the Coulomb gas in the presence of an applied supercurrent.
This more detailed analysis will be useful in later sections when
we analyze Monte Carlo data for the Coulomb gas.

According to the Josephson relation
the voltage $V$ caused by vortex motion is
\begin{displaymath}
V \sim \frac{d \Delta \phi}{d t} \sim n_F I
\end{displaymath}
where we assume that the resistance is proportional to the 
density of ``free'' vortices, $n_F$,
defined by the the Debye-H\"uckel relation.
The linear resistance $R$ is defined by the limit of zero
current $I$:
\begin{equation}
R = \lim_{I \rightarrow 0} \frac{V}{I} \sim n_F
\label{RNF}
\end{equation}
To make an estimate of the density of free vortices we
proceed by the following simple model.
The energy $E(r)$ of a vortex pair of separation $r > r_0$
in the presence of a current $I$ is \cite{AHNS}:
\begin{equation}
E(r) = E_0 + E_1 \ln ( \frac{r}{r_0} ) - I ( r - r_0 )
\label{E0-E1}
\end{equation}
where $E_0$ is a constant, $E_1$ is discussed below and
$r_0$ is the smallest
possible separation, which we will set to $r_0 = 1$ from now.

We will now use the linear screening approximation \cite{minnhagen} to
derive an expression for the second term $E_1\ln r$ in Eq. (\ref{E0-E1}).
The expression is obtained from the Fourier transform of the
linearly screened potential
$V_l(k)$,
\begin{displaymath}
V_l(k)= -{1\over\tilde{\epsilon}(k)}{2\pi\over k^2+\lambda^{-2}}
\end{displaymath}
Here $\lambda$ is the vortex screening length,
and $\tilde{\epsilon}(k)$ is the part of the 
dielectric function, $\epsilon(k)$,
describing the polarization of 
the bound pairs.
The two $\epsilon$ are related by 
\begin{displaymath}
{1\over\epsilon({\bf k})}={1\over\tilde{\epsilon}({\bf k})}
{k^2\over k^2+\lambda^{-2}} 
\end{displaymath}
In the limit $\lambda \rightarrow \infty$ the two different
dielectric functions become equal. This is
the case for temperatures below $T_c$.
The dielectric function, $1/\epsilon(k)$,
is obtained from the charge fluctuations below
in Eq. (\ref{epsilondef}).
The real space expression for $E(r)$ is obtained from
\begin{displaymath}
E(r) = \lim_{\lambda\rightarrow\infty}V_l(r)
\end{displaymath}
Where
\begin{displaymath}
V_l({\bf r}) = \int {d{\bf k}\over 2\pi}^2 V_l({\bf k})
e^{i {\bf k} \cdot {\bf r}} .
\end{displaymath}
We can obtain an approximate expression for $V_l(r)$ by
making use of the fact that $\tilde{\epsilon}({\bf k})$
only depends weakly (in most of $k$ space) on ${\bf k}$.
For a given distance $r$, the Fourier integral picks
up its main contribution from the
$k$ values around $2\pi/r$. Hence
\begin{displaymath}
V_l(r)-V_l(r=1)\approx -
{1\over \tilde{\epsilon}(k=2\pi/r)}K_0(r/\lambda).
\end{displaymath}
Here we have subtracted $V_l(r=1)$ in
order to eliminate the creation
energy. $K_0$ denotes a modified Bessel function.
As $\lambda\rightarrow\infty$ this expression reduces to
\begin{equation}
V_l(r)-V_l(r=1) \approx {1\over \epsilon(2\pi/r)}\ln(r/\lambda)
\label{Vl}
\end{equation}
where we use $\epsilon$ instead of $\tilde{\epsilon}$,
as the temperature is below $T_c$.

According to this discussion the
coefficient $E_1$ is given by \cite{olsson}
\begin{equation}
E_1 = {1\over \epsilon(2\pi/r)}.
\label{E1}
\end{equation}
The weak $r$ dependence describes the
effect of the surrounding vortex pairs.
The coefficient $E_0$ contains the
remaining constant terms from Eq. (\ref{Vl}).
In a first approximation we will
neglect the $r$ dependence in $E_1$.

The energy $E(r)$ in Eq. (\ref{E0-E1}) has a maximum at
separation $r^* = E_1/I$ and the
energy needed to separate a vortex pair
to this distance is \cite{kadin} :
\begin{eqnarray}
\Delta E = E(r^*) - E(r = 1) = E_1 \ln (r^*) - I(r^*-1) \\
\Delta E  = E_1 \ln (\frac{E_1}{I}) - E_1 + I
\end{eqnarray}

Let $\Gamma$ denote the thermal production rate
of free vortices.
A vortex vanishes when it collides with an antivortex.
Hence, it appears reasonable to assume an
annihilation rate proportional
to $n_F^2$. This leads to the following rate equation.
\begin{equation}
\dot{n}_F = \Gamma - cn_F^2.
\label{production}
\end{equation}
Where $c$ is a constant.
The steady state condition is $\dot{n}_F=0$ and
hence we have $\Gamma \propto \sqrt{n_F}$.
Assuming that $\Gamma$ is determined by  activation
over the barrier $\Delta E$ we get the following
production rate\cite{AHNS}
\begin{equation}
\Gamma \propto e^{-\frac{\Delta E}{T}}
\label{gamma}
\end{equation}
and hence for the resistance $R$ from Eq. (\ref{RNF}).
\begin{eqnarray*}
R \propto n_F   \propto e^{-\frac{\Delta E}{2T}} \propto  \\
e^{-\frac{1}{2T}(E_1 \ln(\frac{E_1}{I}) - E_1 + I)} \propto \\
\langle\frac{E_1}{I}\rangle^{-\frac{E_1}{2T}} \
e^{\frac{E_1}{2T}} e^{-\frac{I}{2T}}  
\end{eqnarray*}
keeping the important term for small but finite $I$ we arrive at:
\begin{equation}
R \propto \langle\frac{E_1}{I}\rangle^{-\frac{E_1}{2T}}
\label{currres}
\end{equation}
A given current $I$ gives rise to a ``current
length scale'' $r^*$ from the maximum condition in Eq. (\ref{E0-E1}).
As the lattice of the system has a finite size,
this sets an upper limit to the ``current length'' and hence a
lower limit to the current producing nonlinear resistance.
The smallest current giving nonlinear resistance is $I^*=E_1/r^*$
with $r^* = L$ and hence
for currents smaller than $I^*$ the resistance will be cut
off by the finite size $L$ of the lattice and the resistance
becomes ohmic.
The Nyquist resistance is calculated with $I=0$ and hence
\begin{equation}
R \propto \langle\frac{1}{L}\rangle^{-\frac{E_1}{2T}}.
\label{finite-L}
\end{equation}

This means that we can scale the linear resistance $R$
from the Nyquist relation Eq. (\ref{Nyquist}) with the
exponent $E_1/2T$.
This exponent is precisely $a(T)$, the exponent of
the nonlinear IV characteristics (see Eq. (\ref{IV}) below), hence:
\begin{equation}
f(T) = R L^{a(T)}
\label{RLA}
\end{equation}
should collapse onto a single curve for different lattice sizes $L$.
I.e.\ $f(T)$ should not depend on lattice size $L$.
The resistance we use for this scaling will be the one
determined from the voltage fluctuations Eq. (\ref{Nyquist}).
The exponent determined from resistance data at zero current
will be denoted $a_R(T)$.

\subsection{Nonlinear IV exponent}

We are going to make use of a couple of different expressions for the
power law exponent $a(T)$ of the nonlinear IV characteristics.
{} From Eq. (\ref{currres}) we get the nonlinear
IV characteristic:
\begin{equation}
V \propto \langle\frac{E_1}{I}\rangle^{-\frac{E_1}{2T}} I
\propto I^{a(T) + 1}
\label{IV}
\end{equation}
The exponent calculated by monitoring the voltage
response $V$ as a function of 
an applied supercurrent $I$ will be denoted $a_{IV}(T)$.
On a finite system we will obtain a nonlinear voltage response only
above a finite applied current, given by $I^* \sim E_1 / L$, such that the 
current length $r^*$ is shorter than the size $L$ of the system,
as discussed above.

\subsection{Self consistent IV characteristic}

Another expression for the IV characteristic is obtained if we
include the $r$ dependence in $E_1$ in Eq. (\ref{E1}).
The length dependence can in a first approximation
(in an expansion in derivatives of $E_1(r)$)
be included simply by replacing $E_1$ in Eq. (\ref{currres})
by $1/\epsilon(2\pi/r^*)$ in
the extremum equation $I=E_1/r^*$.
Our rationale for this choice is that at the separation
$r^*$ the vortex pair is broken apart and we therefore
use the stiffness $1/\epsilon(r)$ of the system at this
separation.
We find the appropriate $\epsilon(r)$ by solving
self consistently the equation
\begin{equation}
I=\frac{1}{\epsilon(2\pi/r^*)r^*} =
\frac{k^*}{\epsilon(k^*)2\pi}
\label{sceq}
\end{equation}
The self consistent $\epsilon$
obtained by solving Eq. (\ref{sceq})
will be denoted $\epsilon^*$.
The relation between the exponent $a(T)$ and the
dielectric function $\tilde{\epsilon}$
is according to Eqs. (\ref{currres}) and (\ref{E1})
given by the expression (see Ambegoakar et al. \cite{AHNS})
\begin{equation}
a(T)_{AHNS} = \frac{1}{2 T \epsilon^*}
\label{a-AHNS}
\end{equation}
here we use $\epsilon^*$ as we are at temperatures
below $T_c$.

Recently Minnhagen et al.\ have used scaling arguments to derive
an alternative relationship between $a(T)$ and $\epsilon$,
given by \cite{PM}
\begin{equation}
a(T)_{PM} = \frac{1}{T \epsilon^*} - 2
\label{a-PM}
\end{equation}
As one immediately realises Eq. (\ref{a-PM}) is not consistent
with the activation argument used to derive Eq. (\ref{a-AHNS}).
In order to reconcile Eq. (\ref{a-PM}) with a rate equation like
Eq. (\ref{production}) Minnhagen et al. have made the following suggestion.
They assume that the activation is correctly represented by $\Gamma$ in
Eq. (\ref{gamma}). The recombination, which in Eq. (\ref{production})
is represented by the innocently looking term $n_F^2$, is on the other
hand supposed to be  replaced by $n_F^{1+b}$ with $b=2/(E_1/T-2)$. The
sole argument for this replacement is unfortunately so far simply the
observation that one then can derive Eq. (\ref{a-PM}) from an equation
like Eq. (\ref{production}). Nonetheless, we shall see below that for
temperatures below $T_c$ Eq. (\ref{a-PM}) fits the simulation data much
better than Eq. (\ref{a-AHNS}) does.
However a motivation for a recombination term different
from the one in Eq. (\ref{production}) has not been presented.
At $T_c$ both relations reproduce the same exponent $a(T=T_c)=2$.

\section{Monte Carlo Simulation}

In this section we describe how we calculate 
current-voltage characteristics by Monte Carlo 
simulation of the lattice Coulomb gas.

The algorithm to simulate the lattice Coulomb gas 
works as follows \cite{teitel2}:
First we pick a nearest-neighbor pair $(i,j)$ of lattice sites at random.
Then we try to increase $n_i$ by one and to decrease $n_j$ by one,
thus preserving overall vortex neutrality, $\sum_i n_i = 0$.  
This Monte Carlo move of inserting a neutral pair will be interpreted as
transfer of one unit vortex from site $j$ to $i$.
If the energy change is $\Delta E$ we accept this trial move 
according to the standard Metropolis algorithm \cite{metrop} with 
probability $\exp(-\Delta E/T)$.  
These simple Monte Carlo moves can both
create, annihilate, and move vortices.
Thermodynamic averages are computed 
as Monte Carlo time averages over the sequence of generated configurations.

To calculate IV characteristics works as follows \cite{MWSMG,Hyman}:
An applied current density $j$ gives a Lorentz force
of $j h/(2e)$ on a unit vortex.  The Lorentz force can be incorporated in 
the Monte Carlo moves \cite{MWSMG} by adding to $\Delta E$ an extra term
$j h/(2e)$ if the unit vortex moves in the direction opposite to the Lorentz
force, subtracting this term if it moves in the same direction, 
and making no change in $\Delta E$ if it moves in a perpendicular direction.
Biasing the Monte Carlo moves in this way takes
the system out of equilibrium and
causes a net flux of vortices in a direction perpendicular to the current.
This generates a voltage given by the Josephson relation:
\begin{equation}
V = \frac{h}{2e}  \langle I_v(t) \rangle  , 
\end{equation}
where $I_v(t)$ is the vortex current.
Here $t$ denotes Monte Carlo time, incremented by
$\delta t$ after each attempted move.  The vortex current is
$I_v(t) = +1/L \Delta t$ if a unit vortex moves one lattice spacing in
the direction of the Lorentz force at time $t$, 
and $I_v(t) = 0$ otherwise.
We use units such that $\Delta t = 1/L^2$
so that an attempt is made to move a vortex on each lattice site,
on average, per unit time.
We also use units such that $h/(2e) = 1$.

The linear resistance can also be obtained from the Nyquist relation
in Eq. (\ref{Nyquist})
for equilibrium voltage fluctuations in the absence of any net currents.
For discrete Monte Carlo time it is given by \cite{Reif,APY}, in our units,
\begin{equation}
R = \frac{1}{2T} \sum_{t = -\tau}^{\tau}
\Delta t \langle V(t) V(0) \rangle .
\end{equation}
The cutoff time $\tau$ is set to  1000 time steps, this has
proved to be sufficient as $R$ saturates for values of $\tau$ less
than 100 time steps.

The underlying assumption in using Monte Carlo dynamics to calculate 
IV characteristics is that Monte Carlo time can be equated with 
real time. This approximation has proven reasonable in other 
simulations of vortex dynamics \cite{MWSMG,Hyman,teitel3}.
It should be good near a critical point where vortex motion
is slow and overdamped, but not so satisfactory 
at high temperatures or currents where the discreteness of Monte Carlo time
becomes visible as a saturation of vortex velocities.
We get strong support for this assumption from the results in the next 
section since we can reproduce
the expected IV characteristic at the Kosterlitz-Thouless transition,
and since we come close to experiments.
One can in principle also test this by comparison to 
dynamics simulations where time evolution equations are
integrated \cite{Meakin}.  

In the equilibrium simulations in the case of no net currents
we typically use $10^6-10^7$ Monte Carlo sweeps (one sweep means
one Monte Carlo time step defined above, i.e.\ $L \times L$ attempts to
insert nearest-neighbor pairs),
and in the nonequilibrium case of an applied current
we typically use $10^5-10^6$ Monte Carlo sweeps.
In the evaluation of the Nyquist formula for the linear resistance,
Eq.~(\ref{Nyquist}), we typically sum over $10^3-10^4$ time steps.

The first task for our simulations is to locate the
Kosterlitz-Thouless transition temperature
$T_c$.
The usual universal jump criterion for a
Kosterlitz-Thouless transition involves the
dielectric response function $1/\epsilon$,  given by
\begin{eqnarray}
\frac{1}{\epsilon({\bf k})} &=& 1 - \frac{2 \pi}{k^2 T L^2}
\langle n_{\bf k} n_{\bf -k} \rangle \\
\label{epsilondef}
n_{\bf k} &=& \sum_{{\bf r}_i} n_i e^{-i {\bf k \cdot r}_i},
\end{eqnarray}
where $n_{\bf k}$ is the Fourier transform of the vortex density. 
The limit ${\bf k} \rightarrow 0$ denoted
$1 / \epsilon({\bf k}=0)$, corresponds to the 
fully renormalized long wavelength superfluid density,
and the universal jump criterion tells us that $1 / \epsilon({\bf k}=0)$
jumps from $4T_c$ at $T=T_c^-$ to $0$ at $T=T_c^+$ \cite{NK}.
A practical difficulty for locating $T_c$ from
Monte Carlo data on small lattices 
with this procedure is
that extrapolation to the ${\bf k}=0$ limit requires large lattices,
as the smallest nonzero ${\bf k}$ is $2\pi/L$.
The corresponding quantity to $1 / \epsilon({\bf k}=0)$
in the two dimensional XY model
is called the helicity modulus $\gamma$ \cite{teitel4}.
Both quantities have been used to locate the
Kosterlitz-Thouless transition temperature in Monte Carlo calculations
\cite{teitel4,teitel1,weber}.

In the data analysis in the next section we use an alternative procedure
to locate $T_c$ from the linear resistance \cite{wallweb}.
We obtain the linear resistance $R$ 
from the Nyquist formula in Eq.~(\ref{Nyquist}) for a sequence of 
system sizes $L$ and temperatures $T$.
According to Eq.~(\ref{LogCorrection}) data for $L^2 R$
for different system sizes
should become system size independent at the critical temperature,
which is our criterion to locate $T_c$.
This circumvents the difficulties of using $1 / \epsilon({\bf k}=0)$.
The two determinations give within error bars the same value for $T_c$.

\section{Results}

We present Monte Carlo simulation results for three different
determinations of the IV exponent $a(T)$ for the 2D Coulomb gas model.
The results we show here are for the chemical potential $\mu=0.0$
and lattice size $L = 32$ if not differently stated.

\subsection{Nonlinear IV exponent}

Our first method consists in a direct measurement
of the electric field
$E$ induced by an applied current density  $j$.
In Fig. \ref{IV-CG}a results for the IV  characteristic of the
two-dimensional Coulomb gas are shown.
The dashed line in the $\ln(E)$ versus $\ln(j)$ plot has slope three
and represents the slope at $T=T_c$ according to the universal
jump condition \cite{NK}.
The solid curves represent results for different
temperatures. For very high current the voltage response
saturates. 
This is because when all attempts to move the vortices in
the direction of the Lorentz force are already accepted,
further increasing the current can not give more voltage.
For low enough current there is a crossover to ohmic resistance,
when the current length equals the system size,
and the nonlinear dependence of the resistance on the current vanishes.
The regime where we probe the non-linear
IV characteristic is for this figure
approximately from $\ln j \approx -1.5$ up to
$\approx -0.5$.
According to the Kosterlitz-Thouless  theory the slope of the lines
should be 3 at the critical temperature, and this
criteria can be used to determine $T_c$. We will
however use an independent determination \cite{wallweb}
of $T_c$ for this system, based on the finite size scaling
relation Eq. (\ref{LogCorrection}).

In Fig. \ref{IV-CG}b we demonstrate the
effects of the finite lattice size for low driving 
currents at temperatures below $T_c$.
The data shown are for $T=0.15$ and lattice sizes 
are $L=8$ (triangles), 12 (open squares), 16 (stars),
24 (open circles), and 32 (filled circles).

The finite size effects for the lower temperatures
can be understood in the following way.
(See the discussion above following Eq. (\ref{currres}).)
The finite lattice size is important because pair excitation over
the barrier given by the periodicity length $L$ will add to the
dissipation due to unbinding of pairs over the barrier given by
the pair size $r^*$.
The induced electric field will accordingly be of the form
\begin{equation}
E = R(L)j+constant \ j^{E_1/2T+1},
\label{finite-size}
\end{equation}
where the first term $R(L)$ follows from Eq. (\ref{finite-L})
and $R(L)\rightarrow 0$ as $L\rightarrow\infty$. The second term
in Eq. (\ref{finite-size}) is given by Eq. (\ref{currres}) and will
remain finite in the limit $L\rightarrow \infty$.
This is clearly demonstrated in Fig. \ref{IV-CG}b where
we see that the crossover in  Eq. (\ref{finite-size}) between
the linear and nonlinear regime appears at a
higher driving current for the smaller 
$8 \times 8$ lattice
as the current length ($E_1/I = \xi_I \sim    L$)
associated with the current density $j$
exceeds the size of the lattice.

In Fig. \ref{aX} the exponent $a_{IV}(X)$ is shown as a function
of the reduced temperature $X=T/T_c$.
The dashed horizontal line represents the
universal jump condition for $a_{IV}(X)$.
The plusses represent experimental data from
a superconducting Hg-Xe film \cite{kadin,minnhagenaX}.
The filled circles are the results for
$a_{IV}(T)$ from Fig. \ref{IV-CG}.
The other three data sets are for lattice sizes
$L=16$ (stars), 24 (open circles), and 48 (triangles).
As one can see there are no apparent finite size effects in
the data.
In the vicinity of the critical temperature  the experimental
data are reproduced by the Monte Carlo simulations.

The reduced temperature variable $X$ used in Fig.~\ref{aX} for
the experiment is also from Ref. \cite{minnhagenaX} and for the
lattice Coulomb gas data we use $T_c=0.218$ \ \cite{wallweb}
determined from a finite size scaling analysis using Eq.~(\ref{LogCorrection}).

The inset in Fig. \ref{aX} shows a selection of 
experimental data analyzed along
the lines described 
in Ref.~\cite{minnhagenaX}.
The data in the inset (plusses) are the same
as in the main figure, the other data 
are for Bi$_2$Sr$_2$CaCu$_2$O$_x$
single crystal (filled squares) \cite{gorlova},
and for Bi$_{1.6}$Pb$_{0.4}$Sr$_2$Ca$_2$Cu$_3$O$_x$
single crystal (open squares) \cite{pradhan}.

The Monte Carlo data presented here for $a(X)$ are all for
$\mu=0.0$. We also did the same analysis for Monte Carlo data for
$L=32$ and $\mu = -0.4, -0.2$ and $0.2$.  The closest fit to the
experimental results is produced by $\mu=0.0$. 
Results for different $\mu$ differ from the $\mu=0.0$ results,
by that $\mu=0.2$ has a slightly larger derivative at $X=1$
and the smaller $\mu$ are correspondingly less steep.

\subsection{Self consistent IV characteristic}

In our second determination of the exponent $a(T)$ we will
make use of the relations between $\epsilon^*$
and $a(T)$ in
equations (\ref{a-AHNS}-\ref{a-PM}).
The analysis is based on the self consistent solution of
Eq. (\ref{sceq}). For a given current density $j$, a set of
$\epsilon(k)$ will be calculated for different temperatures.
The self consistent solution to Eq. (\ref{sceq}) for $\epsilon^*$ is shown
in Fig.  \ref{scs}, in (a) data for $T=0.18$ is shown and
in (b) $T=0.24$.
The solid and dashed straight lines represent
$\epsilon^* = k^*/j 2\pi$, given by Eq. (\ref{sceq}),
for different current densities $j$.
The open circles represent $\epsilon({\bf k}^*)$ as a function of
$k$ for different current densities.
The choice of the direction along which $\epsilon({\bf k}^*)$ is
probed is perpendicular to the current density $j$, ie.
parallel to the vortex
drift caused by the current density $j$.
The intersection between a $\epsilon({\bf k}^*)$ curve and
the corresponding straight line is the solution to
Eq. (\ref{sceq}), these are marked with filled circles.
In Fig.  \ref{scs}a for $T=0.18 < T_c=0.218$ we
see that the self consistent solution depends only weakly 
on the choice of probing current as long as the current is not too large.
In Fig. \ref{scs}b for $T=0.24 > T_c$
we see, however, that there is no well defined limiting solution
for $\epsilon^*$ as $j \rightarrow 0$. This is because the
system is above the Kosterlitz-Thouless temperature and vortex
pairs will always dissociate irrespective of $j$.
In Fig.  \ref{scseps} the function $\epsilon^*$ is shown as
a function of temperature.
The solid circles represent $\epsilon^*$ from the self consistent
Eq. (\ref{sceq}) for the fixed current density $j = 0.03125$.
The data shown here represents the construction shown
in Fig. \ref{scs}.

The results from the self consistent solution for $\epsilon^*$ 
are analyzed in Fig. \ref{scsaT}.
Here the filled circles represent the exponent
$a_{IV}(T)$ from Fig. \ref{aX}.
The upside down triangles represent
$a_{AHNS}(T)$ from Eq. (\ref{a-AHNS})
with the solution from Fig. \ref{scs} and the triangles
are the corresponding solution to Eq. (\ref{a-PM}).
One can clearly see that the expression in Eq. (\ref{a-PM}),
derived by Minnhagen et al. \cite{PM},
reproduces the exponent $a_{IV}(T)$ for  $T<T_c$.
Note however, it is only a coincidence that 
Eq. (\ref{a-AHNS}), derived by Ambegoakar et al. \cite{AHNS},
works for temperatures above $T_c$ in this figure as
the limiting ($j \rightarrow 0$)
solution for $1/\epsilon^*$ is not well
defined for these temperatures, as already discussed
in connection with Fig. \ref{scs}b.
As the simulation data $a_{IV}(T)$ (filled circles) also
matched the experimental data in
Fig. \ref{IV-CG} we must conclude that
below $T_c$ the interpretation according to
Eq. (\ref{a-PM}) is clearly
the more appropriate.

\subsection{Linear Resistance}

We will now turn to our last determination of the exponent $a(T)$.
The results presented above all relied on non-equilibrium
Monte Carlo simulations, i.e. with a finite applied supercurrent density $j$.
We will now present the equilibrium determination for $j=0$ based
on finite size scaling of Monte Carlo data for the linear resistance given by the
Nyquist formula (\ref{Nyquist}) together with Eq. (\ref{RLA}).
In Fig. \ref{resscal} we demonstrate a
data collapse of the linear resistance for several
lattice sizes.
{} From Eq. (\ref{RLA}) we see that
the linear resistance data
can be collapsed onto
a single curve, thus representing the thermodynamic limit,
by an appropriate choice at each temperature
$T$ of the exponent $a_R(T)$.
We do this in the following way.  For a given temperature we find
the exponent $a_R$ which minimizes the error of the fit defined as
$\sum_{L,L'} (R(L)L^a - R(L')L'^a)^2$. The considered lattice sizes
are $L,L'=6, 8, 12, 16, 24, 32$. The obtained scaling exponent $a_R$
as a function of temperature is shown in Fig. \ref{resaX}.
As a comparison
we also show  data for $a_{IV}(T)$ from Fig. \ref{aX},
obtained from direct evaluation of the IV characteristic.
The finite size scaling analysis  in
Fig. \ref{resscal} breaks down for low temperatures.
This can be seen by the deviation of  $a_R(T)$ from the
data for  $a_{IV}(T)$ at $T=0.15$.
In  Fig. \ref{resaX} this deviation is also evident.
A careful inspection of the scaling at temperatures
$T=0.15$ and $T=0.18$ reveals that the order
of the lattices sizes is reversed for $T=0.15$ compared with
the higher temperatures. 
This may be related to the difficulties to converge the simulation at
low temperature.

\section{Discussion}

We have calculated the nonlinear IV exponent $a_{IV}(T)$ of
the two dimensional lattice Coulomb gas.
Our results are based on three different determinations.
A direct calculation of the voltage response as a function of
an applied current. Comparison with experiments
\cite{kadin,gorlova,pradhan,minnhagenaX}
on Hg-Xe films and single crystal high $T_c$ superconductors 
show good agreement.

Our second method is based on a simple self consistent
calculation of the dielectric function $\epsilon^*$ at the
unbinding separation, and the IV exponent can then be calculated.
Here we especially focus on the comparison of two relations
between $a(T)$ and $\epsilon$. The first relation
Eq. (\ref{a-AHNS}) \cite{AHNS} is based on
ordinary diffusion in two dimensions with a recombination
rate proportional to $n_F^2$. The second expression for
$a(T)$ given in Eq. (\ref{a-PM}) has been derived from
a scaling analysis \cite{PM}.

We find that the exponent determined by Eq. (\ref{a-PM}) for
temperatures below $T_c$ is close to the more direct
determined $a_{IV}(T)$ and will therefore also fit the
experiments for these temperatures.

The third method is based on equilibrium Monte Carlo
simulations. From the scaling relation Eq. (\ref{RLA}) for the
linear resistance we can derive $a_R(T)$.
We find that the scaling exponent $a_R(T)$ to a high degree
of accuracy fits the direct determined $a_{IV}(T)$ for a
broad range of temperatures.
This provides a link between the
equilibrium and nonequilibrium response properties of the system:
A finite mesoscopic linear resistance in a finite sample 
below $T_c$ is due to
thermally activated vortex motion across some potential barrier,
generated by interactions with all other vortices in the system.
When a finite current is imposed across the system, new nonequilibrium
configurations are accessed where vortices are driven away
from their equilibrium positions by the finite Lorentz force,
thus giving nonlinear response.
Our data shows that the barrier overcome by the Lorentz force,
giving nonlinear response, is essentially the same barrier as in the
equilibrium case, i.e., the potential barrier in the 
case of a finite current appears to be determined by equilibrium 
states in the system.
This is consistent with the scaling ansatz,
discussed above, of a certain
current length scale ($r^*$) 
associated with the finite current, such that 
for lengths shorter than the current length scale
an equilibrium state is still attained, which
gives a potential barrier essentially
equal to that in the equilibrium case.

An interesting possibility arises here to measure
finite size effects on the linear resistance  in 
lithographic Josephson junction arrays. 
The idea would here to take advantage of 
finite size roundings, rather than as usual want 
them to be as small as possible,
and trying to fit experimental data on very small arrays to our finite size 
scaling formulas.  This would provide an unusual 
experimental test of finite size scaling. 
It is also important to analyze the data in terms of the 
reduced temperature scale, as $T_c$ is sample dependent.
According to Eq. (\ref{RLA}) it should be possible
to scale the linear resistance of samples of 
different sizes onto a single scaling function 
using the exponent $a(X)$ from the nonlinear IV characteristics.

Our conclusions are:
(1) Simulation of nonequilibrium vortex dynamics allow
calculation of a lattice size independent IV exponent $a(T)$ as
a function of temperature $T$.
(2) This curve for $a(T)$ agrees nicely with experiments
in an interval around $T_c$,
and this appears to be reported here for the first time.
(3) This curve can be obtained from a simple phenomenological theory
for the nonlinear IV characteristic. 
(4) This curve can also be obtained from a simulation of the 
equilibrium vortex dynamics.
This provides a useful link between driven diffusion and 
equilibrium dynamics of two-dimensional vortex systems.

We acknowledge stimulating 
discussions with P.~Minnhagen and K.~Holmlund.
H.~W. was supported by grants from Carl Trygger, M.~W. was 
supported by grants from the Swedish Natural Science 
Research Council (NFR) and H.~J.~J. was supported by the British EPSRC.

\pagebreak

\begin{figure}
\caption{
Monte Carlo results for the nonlinear IV characteristic
of the two dimensional lattice Coulomb gas.
$E$ is the electrical field 
and $j$ is the supercurrent density.
In (a) data shown are for parameters $L = 32$ and $\mu = 0.0$.
Different curves are for different temperatures,
starting from below $T=0.12, .15, .16, .18, .20, .22, .23,
.24, .26$, and $.30$.
The dashed line has slope 3 which represents the 
Kosterlitz-Thouless transition.
In (b) finite size effects 
according to Eq. (\protect\ref{finite-size}) are demonstrated.
Data here is for $T=0.15$ and lattice sizes
are $L=8$ (triangles), 12 (open squares), 16 (stars),
24 (open circles), and 32 (filled circles).
The dashed line has slope 1 which represents the linear
ohmic regime.
}
\label{IV-CG}
\end{figure}

\begin{figure}
\caption{
Comparison of the exponent $a(X)$ between experiments,
marked with plusses \protect\cite{minnhagenaX},
and Monte Carlo results for the lattice Coulomb gas 
Sizes are $L=16$ (stars), 24 (open circles),
32 (filled circles), and 48 (triangles).
$X=T/T_c$ is the reduced Coulomb gas temperature.
The dashed line corresponds to the universal jump condition for
the exponent, $a(X) = 2$.
The inset shows experimental data for
Hg-Xe alloy films (plusses) \protect\cite{kadin,minnhagenaX}, 
Bi$_2$Sr$_2$CaCu$_2$O$_x$ single crystal films
(filled squares) \protect\cite{gorlova}, and
Bi$_{1.6}$Pb$_{0.4}$Sr$_2$Ca$_2$Cu$_3$O$_x$
single crystal films (open squares) \protect\cite{pradhan}.
}
\label{aX}
\end{figure}

\begin{figure}
\caption{
Construction of the self consistent solution. The data
shown is for $T=0.18$ in (a) and $T=0.240$ in (b).
The different curves represent  $\epsilon(k)$ as
a function of $k_x$ for different current densities 
$j = 0.05$ (bottom curve), $0.10, 0.15, 0.20, 0.30$ (top curve).
The straight lines represent
the self-consistency condition for the different current densities,
and the intersections of the straight lines with
the  $\epsilon(k)$ curve for the same current density $j$
represents the self consistent solution, here marked with
large filled circles.
}
\label{scs}
\end{figure}

\begin{figure}
\caption{
The solution to the self consistent equation (\protect\ref{sceq}).
$\epsilon^*$ as a function of temperature for the fixed current
density $j = 0.03125$.
Data shown are for size $L=32$ chemical potential $\mu=0.0$.
The dashed line represents the bare $\epsilon = 1$.
}
\label{scseps}
\end{figure}

\begin{figure}
\caption{
The IV exponent $a(T)$ as a function of temperature. Data
shown are for size $L=32$ and chemical potential $\mu=0.0$.
The filled circles are $a_{IV}(T)$ from the
the nonlinear IV characteristic
shown in Fig. \protect\ref{IV-CG}b.
The upside down triangles is the
exponent $a_{AHNS}(T)$ from Eq. (\protect\ref{a-AHNS})
using $\epsilon^*$ from Fig. (\protect\ref{scseps}).
The triangles represent the exponent $a_{PM}(T)$
from Eq. (\protect\ref{a-PM}) also
using $\epsilon^*$ from Fig. \protect\ref{scseps}.
}
\label{scsaT}
\end{figure}

\begin{figure}
\caption{
Finite size scaling of the linear resistance $R$. The resistance 
has been calculated with the Nyquist formula (\protect\ref{Nyquist}).
Shown is a data collapse of $R L^{a_R(T)}$ as a function
of $T$ for different lattice sizes, $L = 6, 8, 12, 16, 24, 32$.
The exponent $a_R(T)$ is chosen in such a way as to
provide the best data collapse.
According to Eq. (\protect\ref{RLA}) the function
$R L^{a_R(T)}$ should be independent of system size.
}
\label{resscal}
\end{figure}

\begin{figure}
\caption{
Comparison of the exponent $a(T)$ as a function of
temperature $T$, from two different determinations.
The filled circles represent the exponent $a_{IV}(T)$
determined in Fig. \protect\ref{aX} from the 
nonlinear IV characteristics. 
The open circles are results for $a_R(T)$ 
from Fig. \protect\ref{resscal} from 
finite size scaling of the linear resistance.
}
\label{resaX}
\end{figure}

\end{document}